# An experimental investigation of adhesive wear extension in fretting interface: application of the contact oxygenation concept

Soha Baydoun, Siegfried Fouvry*

Ecole Centrale de Lyon, LTDS Laboratory, 36 av Guy de Collongue, 69130 Ecully, France

*Corresponding author at: Ecole Centrale de Lyon, LTDS Laboratory, 36 av Guy de Collongue, 69130 Ecully, France

Corresponding authors email addresses: soha.baydoun@ec-lyon.fr (S. Baydoun) and siegfried.fouvry@ec-lyon.fr (S. Fouvry)

**Abstract**

This paper investigates the transition from abrasive to adhesive wear in gross-slip fretting assuming contact oxygenation concept which suggests that adhesion appears in the inner part of the interface if the di-oxygen partial pressure is below a threshold value. In the lateral sides, where di-oxygen molecules are sufficient, oxidation and abrasion prevail. To assess this phenomenon, 34NiCrMo16 flat-on-flat contacts are tested. Contact oxygenation is quantified using the "oxygen-distance, $d_O$" parameter defined as the averaged width of the external abrasion corona. Confirming this concept, $d_O$ decreases with contact pressure and frequency but remains constant versus sliding amplitude, fretting cycles and contact area. $d_O$ evolution is formalized using a power law formulation which allowed predicting wear transitions for plain and macro-textured surfaces.

## 1. Introduction

Gross slip fretting wear was extensively investigated during the past decades underlining synergic interactions between mechanical loadings and tribo-oxidational processes [1–7]. Fretting sliding amplitudes are usually very small compared to the contact size (<10 %), so most of the fretted interface remains hidden from the ambient air. However, oxygen can diffuse within the interface inducing composite adhesive-abrasive wear phenomenon. Hence, for low loading conditions (i.e. low "p.v" factor, with "p" being the pressure and "v" being the sliding speed), pure abrasive wear mechanisms displaying an oxide debris layer over the whole interface are commonly observed. Higher "p.v" conditions promote composite adhesive-abrasive interface displaying metal transfers in the inner part of the contact surrounded by an abrasive wear corona on the lateral sides (Fig. 1). Predicting the adhesive wear phenomena in gross slip fretting wear interfaces is a key issue for many industrial assemblies. Indeed, as depicted by Hintikka et al. [7], such adhesive seizure phenomena may induce local over straining favoring fretting crack nucleation and incipient crack propagation. This typical local over stressing process can explain the dramatic fretting fatigue failures observed in large conformal interfaces like crankshaft, bolted and riveted plates whereas usual fatigue stress analyses are unable to predict any cracking risk due to the very low pressure conditions. Hence, there is a key interest to formalize the transition from "safe cracking" abrasive interface to "dangerous" adhesive-abrasive fretting scar.

Several authors in literature tried to predict transitions in wear mechanisms under different tribological parameters. Lim and co-authors [1,8] proposed a wear-mode map for unlubricated sliding whereby the transitions between mild and severe wear are dictated by the effects of the normal load and the sliding velocity. On the other hand, the transition from adhesive to oxidational wear was shown to be highly dependent on the atmospheric environments [9–11] where adhesion is favored when the oxygen partial pressure is relatively low [12]. Recent approach linking wear transitions to the external environment was suggested by Mary and co-workers [13]. While investigating the influence of contact pressure on titanium alloy, the authors suggested that the significant adhesive wear observed at high pressure is associated to the very poor oxygen supply to the center of the scar. This gives birth to a hard "white layer [14]" commonly referred to as "tribologically transformed structure (TTS)" [15–17] which is harder than the bulk and characterized by low oxide contents. This layer comes from different sources as reported by many authors including frictional heat [14,18], diffusion [19,20], material transfer [15], mechanical mixing [21], and dynamic recrystallization [22]. However, abrasive wear is favored at the edges of the contact due to the easier access of dioxygen molecules from the external ambient air. On the other hand, an abrasion-oxidation process is activated over the whole interface at low pressure because of the easier oxygen penetration to the inner part of the contact. Based on the these findings, Mary et al. [13] introduced the concept of "air distillation". In the same sense, Warmuth and co-authors [23,24] noticed that high radius contacts showed preferential adhesive damage towards their centers due to the oxygen limitation imposed by the contact size. They called this "oxygen exclusion". Besides, the effect of frequency on adhesive wear extension was equally investigated by many researchers including Shipway and co-authors [5,24,25] who revealed that increasing the sliding frequency tends to increase adhesion due to the lower time given for asperity interactions and hence for oxide formation in addition to the higher friction power dissipation [4] which leads to more oxygen consumption resulting in an oxygen-starved environment towards the center of the contact [20]. Studying several loading conditions, Fouvry and co-workers [20] finally combined the preceding notions under a more generalized "contact oxygenation" concept. This latter may be formalized assuming the Interfacial di-Oxygen partial Pressure (IOP) "$P_{O_2}$" parameter. This latter is very high at the contact borders, equivalent to the ambient air, but tends to decrease inside the interface. The decreasing evolution of "$P_{O_2}$" toward the inner part of the contact illustrated by the hyperbolic function in Fig. 1 is related to a balance between the diffusion of dioxygen molecules from the external part of the contact and the consumption rate induced by the oxidation of the fresh surface metal exposed by the friction processes.

The dioxygen diffusion process is controlled by the interface properties such as the clearance between the contact asperities and the diffusion properties of the porous debris layer. The dioxygen consumption rate is governed by the metal reactivity and the surface damage rate which could be qualified using the so-called "p.v" factor. Hence, the higher the "p.v" factor, the faster the degradation of the protective oxide layer, and the higher the fresh metal exposure, the faster the consumption of the available dioxygen molecules and therefore the faster the decreasing of the Interfacial di-Oxygen partial Pressure (IOP) or "$P_{O_2}$".

According to the given "contact oxygenation" concept [20], abrasive-oxidational wear is predominant in the condition of full contact oxygenation. This suggests that whatever the test duration and the contact position inside the fretted interface, the local IOP ($P_{O_2}$) is higher than a threshold value $P_{O_2,th}$ above which the surface oxidation is guaranteed. Alternatively, composite abrasive-adhesive wear process is observed when $P_{O_2}$ falls below this threshold value. Hence, there is a critical "oxygen distance $d_O$" from the contact border above which IOP is no more sufficient so that seizure phenomena are activated (Fig. 1). Assuming "$r$" the radial position from the center of a 3D contact and "$a$" the corresponding contact radius, the transition from a pure abrasive to a composite abrasive-adhesive wear may be expressed considering the following simple relationship:

-pure abrasive wear if whatever $r < a$, $P_{O_2}[r] > P_{O_2,th}$ (1)

-composite adhesive – abrasive wear:

if $r > a - d_O$, $P_{O_2}[r] > P_{O_2,th}$ ⇒ external abrasion zone

if $r < a - d_O$, $P_{O_2}[r] < P_{O_2,th}$ ⇒ inner adhesion zone (2)

To experimentally demonstrate such "contact oxygenation" concept, the given research work investigates the evolution of "$d_O$" (i.e. the lateral width of abrasive zone) as a function of various contact loadings, contact size but also macro-textured surfaces. This is currently achieved thanks to a multi-scale experimental strategy applied on a flat-on-flat contact configuration where oxygen distance evolution can be easily tracked as detailed in Fig. 1. The crossed flat configuration implies an $(X, Y)$ description so that the former axisymmetric description is turned to a conjugate 2D approach:

-pure abrasive wear if whatever $X < L/2$ and $Y < W/2$, $P_{O_2} > P_{O_2,th}$ (3)

-composite adhesive – abrasive wear:

if $X > L/2 - d_O$ or $Y > W/2 - d_O$, $P_{O_2} > P_{O_2,th}$ ⇒ external abrasive wear   (4)

Where $L$ and $W$ are the longitudinal and transverse widths of the studied crossed flat contact interface.

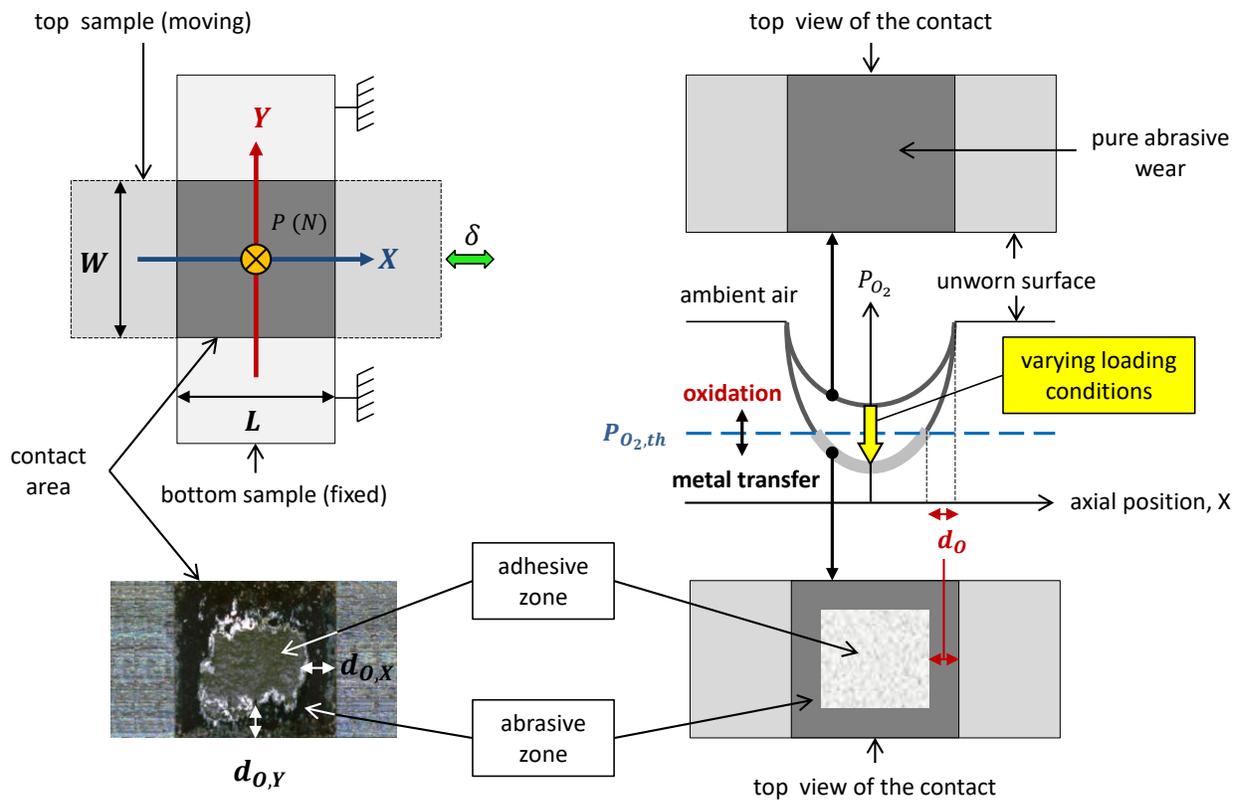

*Fig. 1. Schematic presentation of the transition of wear regimes from pure abrasive wear at high oxygen partial pressure ($P_{O_2} > P_{O_2,th}$) to a mixed abrasive-adhesive wear regime at low oxygen partial pressure ($P_{O_2} < P_{O_2,th}$), After [20].*

## 2. Experimental procedure

2.1 Materials

The samples used are made of a homogeneous tempered low steel alloy (34NiCrMo16) whose mechanical properties are shown in Table 1. Due to its high strength, and hardenability and its good dimensional stability, 34NiCrMo16 is used in a broad range of industrial applications such as aerospace, mechanical parts, and dies.

**Table 1.** *Mechanical properties of 34NiCrMo16 (obtained from the documentation of material supplier [26]).*

| Young's modulus, E (GPa) | Poisson's ratio, v | Yield stress (0.2%), $\sigma_{y\,0.2\%}$ (MPa) | Ultimate stress, $\sigma_u$ (MPa) |
|---|---|---|---|
| 205 | 0.3 | 950 | 1130 |

2.2 Experimental setup and data acquisition

A hydraulic fretting wear test system (Fig. 2a) customized at LTDS (Laboratory of Tribology and Systems Dynamics) to investigate large horizontal crossed flat-on-flat configuration is used. Tests were conducted at ambient conditions of temperature (25 °C ± 5 °C) and humidity (RH=40% ± 10%) such that the bottom sample is fixed, whereas the top sample is displaced thanks to an MTS (MTS Systems Corporation) hydraulic actuator. During the test, the fretting displacement ($\delta$), the tangential force (Q) and the normal force (P) are measured (Fig. 2b). The evolution of "Q" versus "$\delta$" allows plotting the "Q- $\delta$" fretting cycles cumulated during the test giving birth to the so-called fretting log (Fig. 2c). The displacement amplitude is partly a function of the elastic accommodation of the test system which itself depends on the tangential force amplitude. To guarantee a given sliding amplitude, gross slip fretting tests were performed by monitoring the sliding amplitude $\delta_g$, defined as the residual displacement when Q=0. On the other hand, $\delta^*$ is continuously adjusted to keep a constant $\delta_g$ at a given required value.

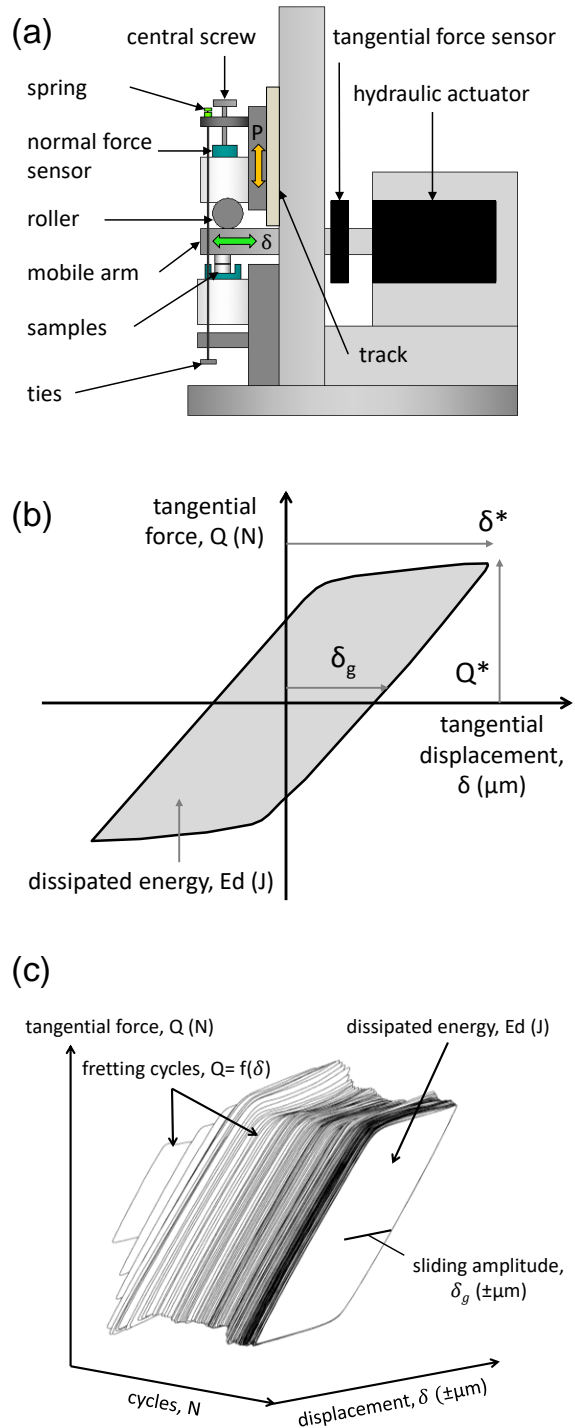

*Fig. 2.* Schematic illustration of the: (a) experimental set-up at LTDS Laboratory; (b) parameters obtained from fretting cycle under gross slip condition; and (c) fretting log with $δ_g$ maintained constant by adjusting $δ^*$.

## 2.3 Contact geometry and experimental strategy

Fig. 3 illustrates the homogeneous crossed flat-on-flat contact configuration used in the current study. This configuration allows a constant contact area and consequently a constant contact pressure during fretting wear test [27]. The fretted interface has a rectangular shape having a contact area "$A = W.L$" where "$L$" is the longitudinal contact size being always parallel to the sliding direction (δ) and "$W$" is the transverse contact size being always oriented against the sliding direction (δ).

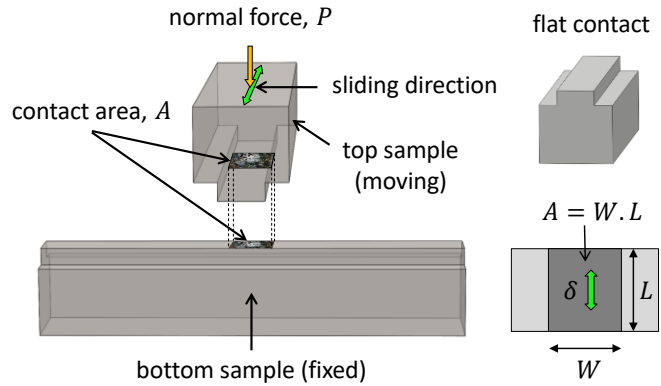

***Fig. 3.*** *Schematic presentation of the crossed flat-on-flat contact configuration.*

The influence of the loading conditions on "$d_o$" parameter is investigated using a multi-scale experimental strategy assuming a key reference test condition (repeated three times) defined by a number of cycles N=20000 cycles, contact pressure p=100 MPa, sliding amplitude $δ_g$= ±100 µm, frequency f=1 Hz, and a square contact area A=25 mm² ($W=L=5$ mm) (Fig. 4). A cross-test strategy is then applied where the contact pressure is varied from p=25 to 175 MPa (Fig. 4a), the sliding amplitude from $δ_g$= ±25 to ±200 µm (Fig. 4a), the frequency from f=0.5 to 10 Hz (Fig. 4b), the number of cycles from N=5000 to 40000 cycles (Fig. 4b), and finally the contact area from A=10 to 25 mm² (Fig. 4c). Note that the contact area is varied firstly by reducing the length "$L$" along the sliding direction from 5 to 2 mm while maintaining a constant length "$W=5$ mm". Secondly, the contact area is varied by reducing the length "$W$" against the sliding direction from 5 to 2 mm while maintaining a constant length "$L=5$ mm". The objective behind this is to evaluate the effect of contact orientation regarding the "$d_o$" parameter.

For such contact configuration the minimum distance from center of the contact to its borders is given by:

$$d = minimum\ of\ \left(\frac{L}{2}, \frac{W}{2}\right) \quad (5)$$

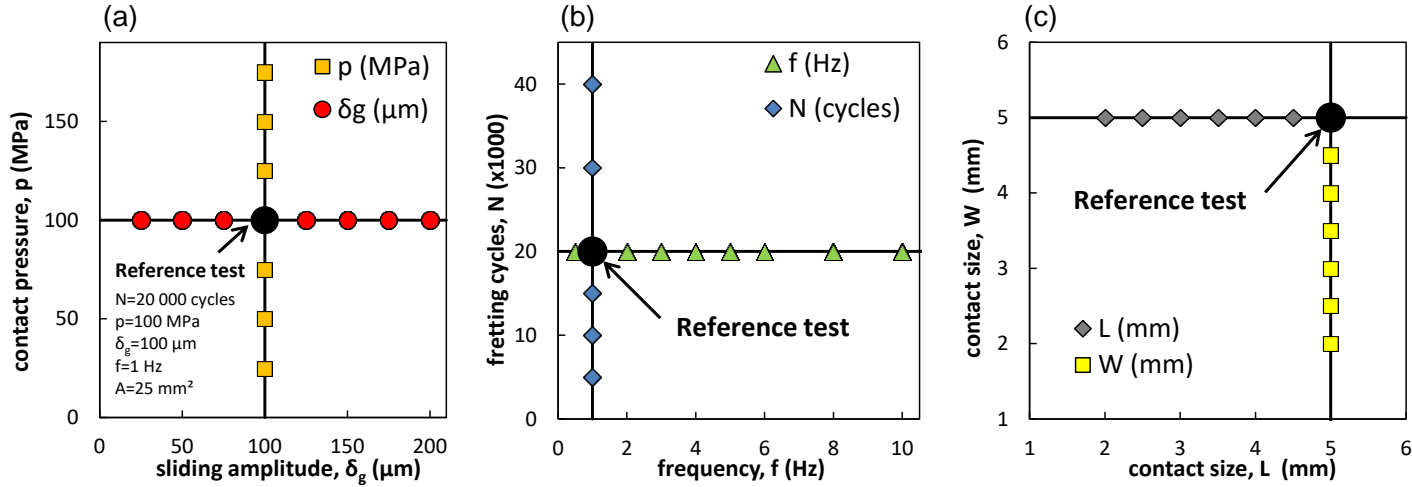

***Fig. 4.** Multi-scale experimental strategy to assess the evolution of oxygen distance "$d_O$": (a) contact pressure versus sliding amplitude; (b) fretting cycles versus frequency; and (c) contact length (W) versus contact length (L).*

2.4 Experimental estimation of oxygen distance "$d_O$" parameter from fretting scar expertise

Top view observations (BSE, EDX) of a mixed abrasive-adhesive wear regime (Fig. 5) show a composite wear scar characterized by a central bright zone and an oxidized corona. Line EDX scans display very low oxide content in the central part compared to the borders. A marked difference in the third body morphology is detected inside and outside adhesive zone. In the central part of the contact where adhesion occurs, the poorly oxidized third body exhibits a bright round to angular shaped metallic protrusions whereas more homogeneous compacted oxide debris is detected in the abrasive zone. This observation is confirmed by cross section view. Thin oxide debris layers having 20 µm thickness were observed on the lateral abrasive wear domain whereas partially transformed TTS structure with a thickness reaching 150 µm was detected in the central adhesive wear zone. The inner TTS structure is characterized by micro-cracks and a very poor oxygen content equivalent to the bulk as reflected by EDX and BSE analyses. Similar observations were detected by various authors for high strength steels [5,28] and for titanium alloys [13,20].

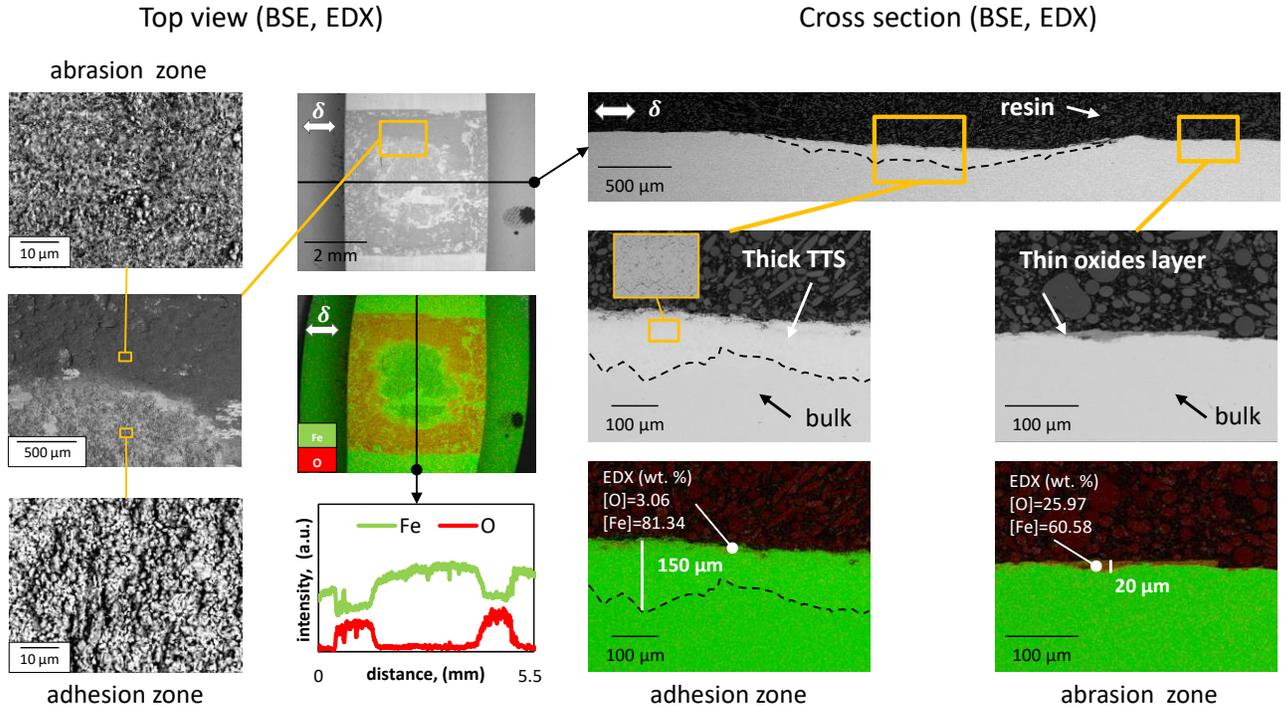

*Fig. 5.* BSE and EDX observations of the top and cross section views of mixed abrasive-adhesive wear scar detected at the bottom sample (N=20000 cycles, p=100 MPa, $δ_g$=±100 μm, f=5 Hz, and A=25 mm²).

A direct estimation of oxygen distance "$d_O$" parameter is not possible. This requires a local in situ analysis of the air composition which is currently technically unfeasible. However, it could be indirectly approximated by measuring the distance between the boundary marking the transition from abrasive to adhesive wear, and the contact borders (Fig. 5). On the other hand, by the measuring the width of the abrasive corona $d_{ab}$ we can indirectly estimate the oxygen distance $d_O$ assuming that:

$$d_O \approx d_{ab} \qquad (6)$$

The abrasive corona also corresponds to the high EDX oxygen concentration which indirectly supports the correlation between $d_{ab}$ and $d_O$. To determine the representative

"$d_O$" value, postmortem EDX oxygen profiles along the longitudinal sliding axis (X) and the transversal axis (Y) were performed (Fig. 6a). Using these two crossed oxygen profiles, the following parameters are estimated:

$$d_{O,X} = \frac{1}{2} \cdot (d_{O,X_1} + d_{O,X_2}) \qquad (7)$$

$$d_{O,Y} = \frac{1}{2} \cdot (d_{O,Y_1} + d_{O,Y_2}) \qquad (8)$$

$$d_O = \frac{1}{2} \cdot (d_{O,X} + d_{O,Y}) \qquad (9)$$

Using the same concept, the adhesion area $A_{ad}$ can be estimated from the regions where the oxygen intensity is relatively low:

$$A_{ad} = d_{ad,X} \cdot d_{ad,Y} \qquad (10)$$

The abrasion area $A_{ab}$ is deduced from the measured adhesion area $A_{ad}$ so that:

$$A_{ab} = A - A_{ad} \qquad (11)$$

The relative proportions of abrasion and adhesion areas are expressed by:

$$\%A_{ab} = \frac{A_{ab}}{A} \text{ and } \%A_{ad} = \frac{A_{ad}}{A} \qquad (12)$$

Fig. 6b plots the evolution of $d_{O,Y}$ as a function of the corresponding $d_{O,X}$ values. The results are quite dispersed which can be explained by the inherent experimental dispersions (small misalignments, dissymmetry of the tangential apparatus compliance, etc. …). To check the symmetry between the oxygen distances $d_{O,X}$ and $d_{O,Y}$ along the longitudinal sliding axis (X) and the transversal axis (Y) respectively, a median line (y=x) is traced. It is interesting to note that the data are symmetrically dispersed along the median curve. This suggests that the extension of the abrasive zone (i.e. well oxygenated interface so that $P_{O_2} > P_{O_2,th}$), is isotropic and non-dependent of the sliding direction and the related debris flow. Assuming such isotropic behavior, $d_O$ which corresponds to the average value of four measurements (Eq. 7-9) along X and Y directions, provides a representative value of the abrasive area extension.

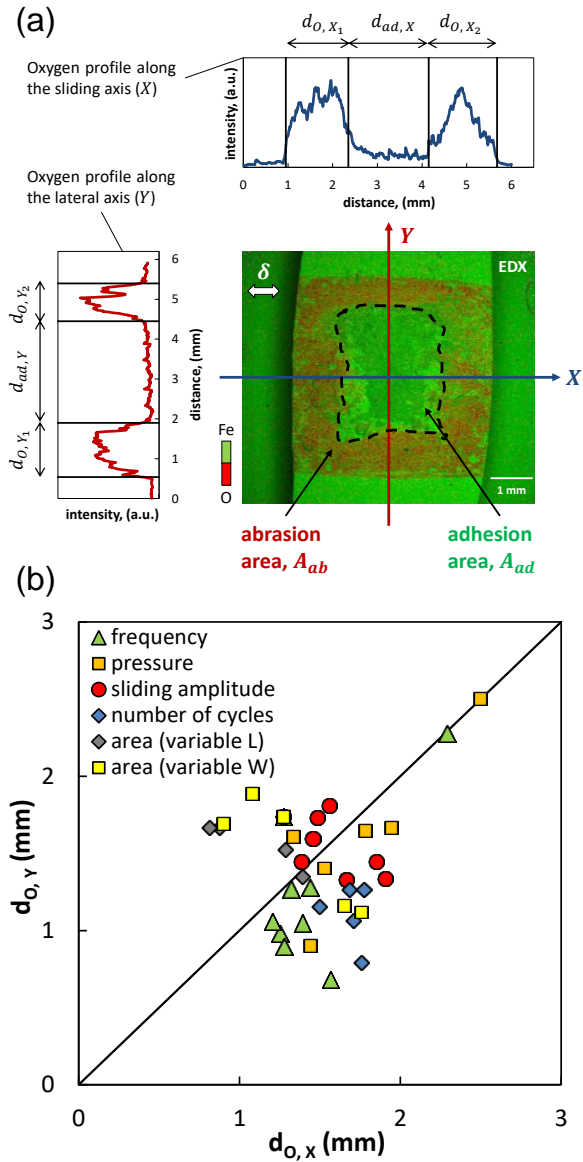

***Fig. 6.*** *(a) Computation of the oxygen distance "$d_O$" and the adhesion and abrasion areas by using crossed EDX oxygen line scans method; (b) plotting of $d_{O,Y}$ versus $d_{O,X}$ by compiling all the loading conditions.*

## 3. Experimental results

This section compiles the different results obtained from plain crossed flat-on-flat contact configuration then from the response of macro textured samples which permit decreasing the mean distance from the inner part of the contact to the external open air allowing an easier investigation of the contact oxygen access.

3.1 Plain crossed flat-on-flat contact configuration

The following section assesses the influence of the studied parameters (Fig. 4) on the relative extension of abrasion and adhesion areas and the related evolution of oxygen distance $d_O$.

3.1.1 Effect of the frequency

Fig. 7a compares wear scars using optical images and EDX maps and line scans for a frequency ranging from 0.5 to 10 Hz. It appears that the inner adhesion area which is negligible at 0.5 Hz extends progressively with the frequency until it becomes significant at 10 Hz. This leads to an asymptotic decrease in the relative proportion of abrasive wear from nearly 100% at 0 Hz until a stabilized value around 69% for the highest frequency values such that $\%A_{ab} = 0.88 \cdot f^{-0.12}$. It is interesting to note that at 10 Hz, the adhesion area displays a well-defined square shape confirming the former hypothesis assuming that the oxygen distance $d_O$ is equal in all directions.

Fig. 7b displays the evolution of the corresponding $d_O$ value defined from Eq. 9. Considering the reference test condition ($f_{ref}$=1 Hz), the evolution of $d_O$ versus the frequency may be expressed using a decreasing power law function so that:

$$d_O = K_f \times d_{O,ref} \times \left(\frac{f}{f_{ref}}\right)^{n_f} \tag{13}$$

With $d_{O,ref}$ =1.51 mm, the oxygen distance related to the reference test condition (N=20000, p=100 MPa, $\delta_g$ =±100 µm, f=1 Hz, and $A = W \times L$ =5 x 5=25 mm²), $n_f$ = -0.22 the exponent marking the decreasing evolution of oxygen distance with frequency and $K_f$=1.12, which is an experimental constant defined by the best fitting of the global evolution of $d_O$ from the chosen reference.

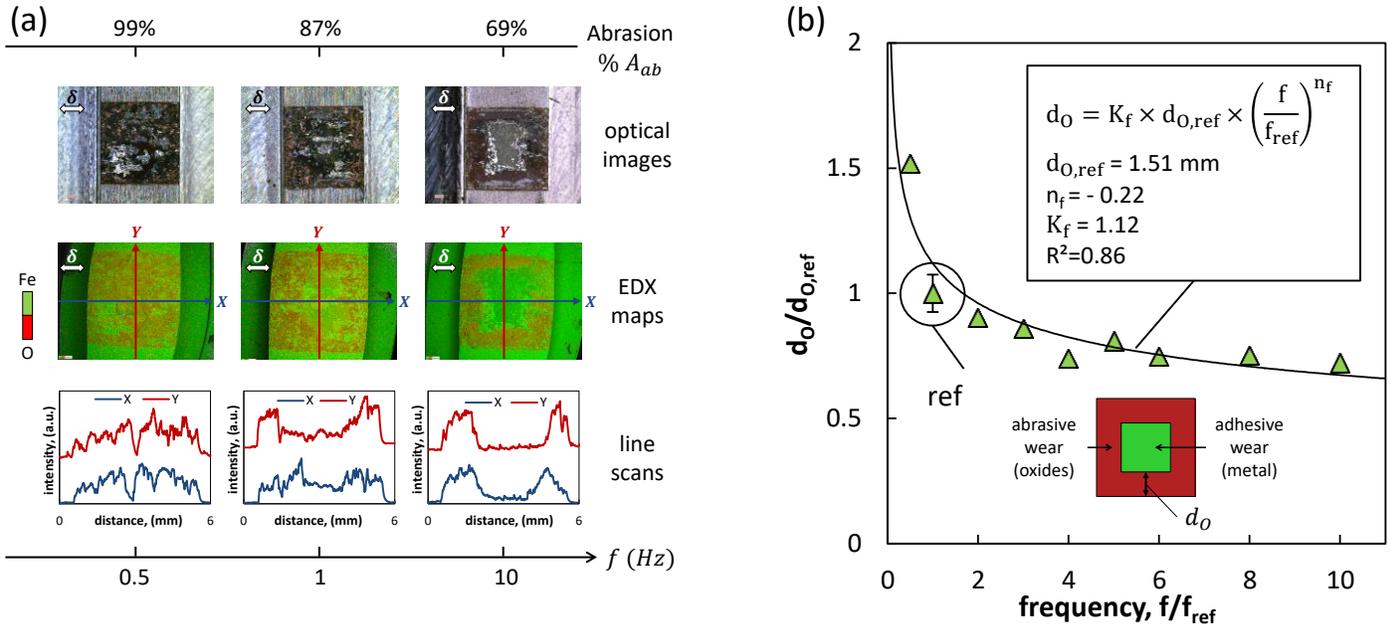

***Fig. 7.*** *(a) Optical images and EDX observations showing the recession of abrasive wear with the increase in frequency from 0.5 to 10 Hz (N=20000 cycles, p=100 MPa, δ$_g$=±100 µm, and A=5x5=25 mm²); (b) Evolution of the oxygen distance "d$_O$" as a function of the frequency (N=20000 cycles, p=100 MPa, δ$_g$=±100 µm, A=5x5=25 mm² with f$_{ref}$ =1 Hz and $d_{O,ref}$ = 1.51 mm).*

### 3.1.2 Effect of the contact pressure

The contact pressure is increased from 25 MPa to 175 MPa to detect its role on fretting wear regimes. Optical observations and EDX maps and line scans (Fig. 8a) reveal a transition from nearly pure abrasive wear at low pressure (p=25 MPa) to a mixed abrasive-adhesive wear response at high pressure (p=175 MPa). This leads to shrinkage of the abrasion zone with the increase in the contact pressure such that $\%A_{ab} = 1.55 \cdot p^{-0.13}$.

This recession of abrasion area is translated in reduction of the oxygen distance by 53% when the contact pressure is increased by 150 MPa as reflected in Fig.8b. Again the evolution of the oxygen distance parameter can be formalized using a simple power law function normalized as a function of the reference test condition so that:

$$d_O = K_p \times d_{O,ref} \times \left(\frac{p}{p_{ref}}\right)^{n_p} \tag{14}$$

With $n_p$ = -0.32 the exponent marking the decreasing evolution of oxygen distance with the contact pressure and $K_p$ = 1.02 the experimental correcting factor related to the best fitting of the global evolution of experimental data.

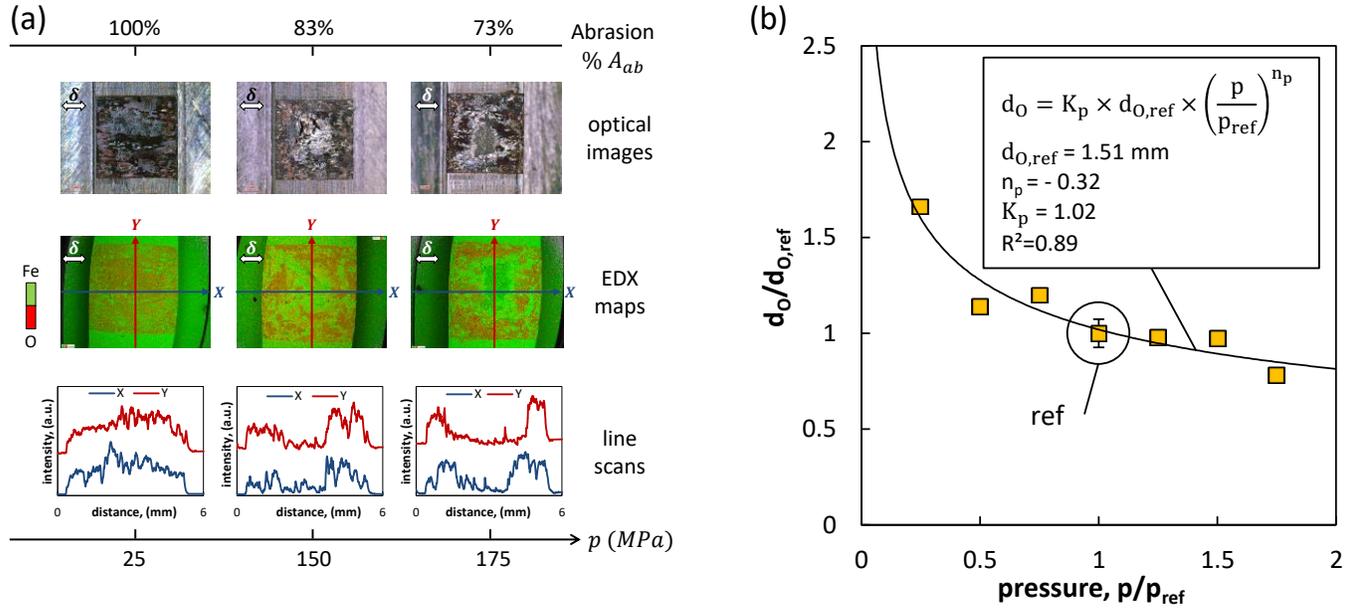

*Fig. 8. (a) Optical images and EDX observations showing the extension of the inner adhesive zone with the increase in contact pressure from 25 to 175 MPa (N=20000 cycles, δ$_g$=±100 μm, f=1 Hz, and A=5x5=25 mm²); (b) Evolution of the oxygen distance "$d_O$" as a function of the mean contact pressure (N=20000 cycles, δ$_g$=±100 μm, f=1 Hz, A=5x5=25 mm² with $p_{ref}$=100 MPa and $d_{O,ref}$ = 1.51 mm).*

### 3.1.3 Effect of the loading cycles

Fig. 9a assesses the effect of the number of cycles by comparing wear scars using optical images and EDX maps and line scans at different test durations. No significant differences are observed where the percentage of the abrasive wear area remains constant (%$A_{ab}$= 0.83 ± 0.03) resulting in a constant $d_O$ value (Figs. 9b) so that:

$$d_O = K_N \times d_{O,ref} \approx 1.42 \; mm \tag{15}$$

With $K_N = 0.94$ being a correcting factor of the mean value compared to the reference test condition ($d_{O,ref} = 1.51$ mm).

This suggests that the stabilized response of the interface is reached before 5000 cycles and the partition between adhesive and abrasive wear remains constant, independent of the test duration at least for the studied contact area conditions.

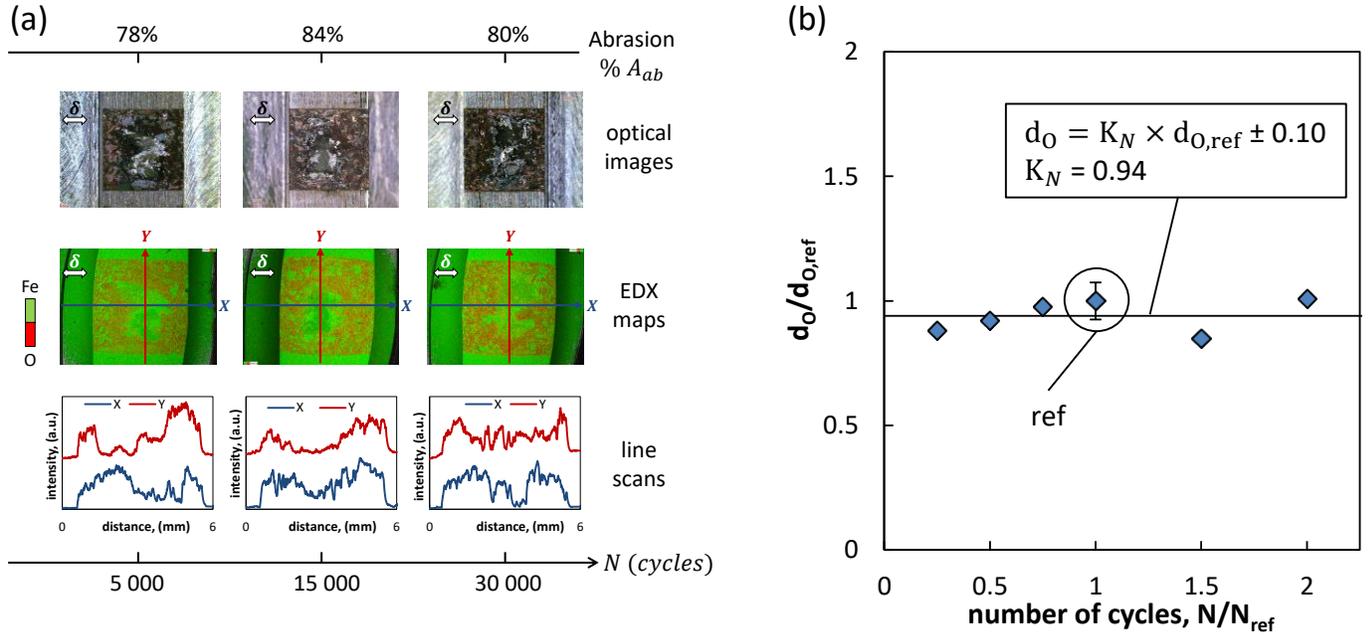

*Fig. 9.* *(a) Optical images and EDX observations showing the evolution of abrasive wear with the increase in number of cycles from 5000 to 30000 cycles (p=100 MPa, δ$_g$=±100 µm, f=1 Hz, and A=5x5=25 mm²); (b) Evolution of the oxygen distance "$d_O$" as a function of fretting cycles (p=100 MPa, δ$_g$=±100 µm, f=1 Hz, and A=5x5=25 mm², with N$_{ref}$ =20000 and $d_{O,ref} = 1.51$ mm).*

### 3.1.4 Effect of the sliding amplitude

The effect of the sliding amplitude on abrasion and adhesion is investigated by comparing optical observations and EDX maps and line scans at different sliding amplitudes (Fig. 10a). No substantial differences are observed. Similar mixed abrasive-adhesive wear partitions are noticed where the percentage of abrasive wear remains constant (%$A_{ab}$= 0.87 ± 0.03) suggesting that the sliding amplitude plays a minor role regarding the dioxygen diffusion processes within the interface.

This is also reflected in Fig. 10b where the oxygen distance stabilizes at $d_O$=1.57 mm whatever the applied sliding amplitude which implies:

$$d_O = K_\delta \times d_{O,ref} \approx 1.57 \, mm \tag{16}$$

With $K_\delta$ = 1.04 being an experimental correcting factor of the mean value related to the sliding amplitude analysis.

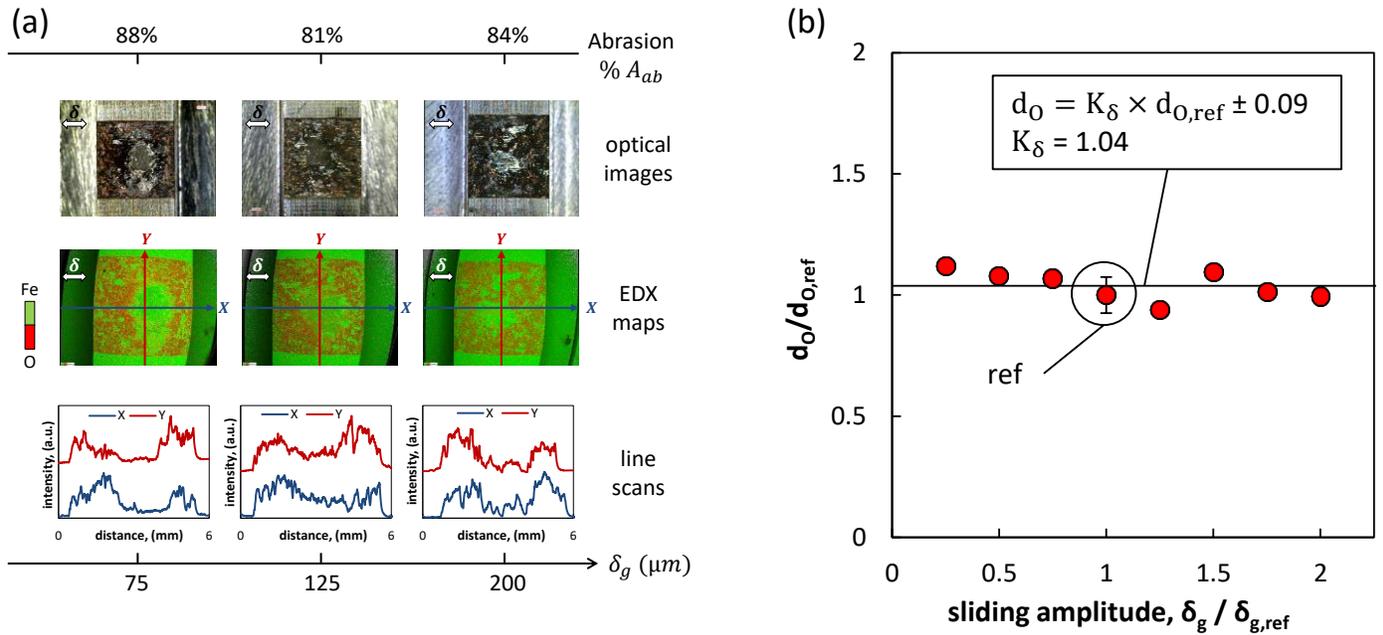

**Fig. 10.** (a) Optical images and EDX observations showing the evolution of abrasive wear with the increase in sliding amplitude from ±75 to ±200 µm (N=20000 cycles, p=100 MPa, f=1 Hz, A=5x5 =25 mm²);(b) Evolution of the oxygen distance "$d_O$" as the function of the sliding amplitude (p=100 MPa, f=1 Hz, and A=5x5=25 mm², with $\delta_{g,ref}$= ±100 µm and $d_{O,ref}$ = 1.51 mm).

3.1.5  Effect of the contact area and the contact orientation

In this section, the evolution of the oxygen distance is investigated as a function of both the contact area and the contact orientation. The contact area is decreased from 25 to 10 mm² (Fig. 3 &4c) by reducing firstly the length "$L$" along the sliding direction from 5 to 2 mm while maintaining constant "$W$=5 mm", then by reducing the length "$W$" against the sliding direction from 5 to 2 mm while maintaining constant "$L$=5 mm".

Decreasing the contact area either by reducing $W$ (Fig. 11a) or $L$ (Fig. 11b) promotes a remarkable regression in the adhesive wear area until it completely vanishes when the contact area is less than or equal to 12.5 mm². This is confirmed by comparing optical images and EDX lines scans and maps for both orientations.

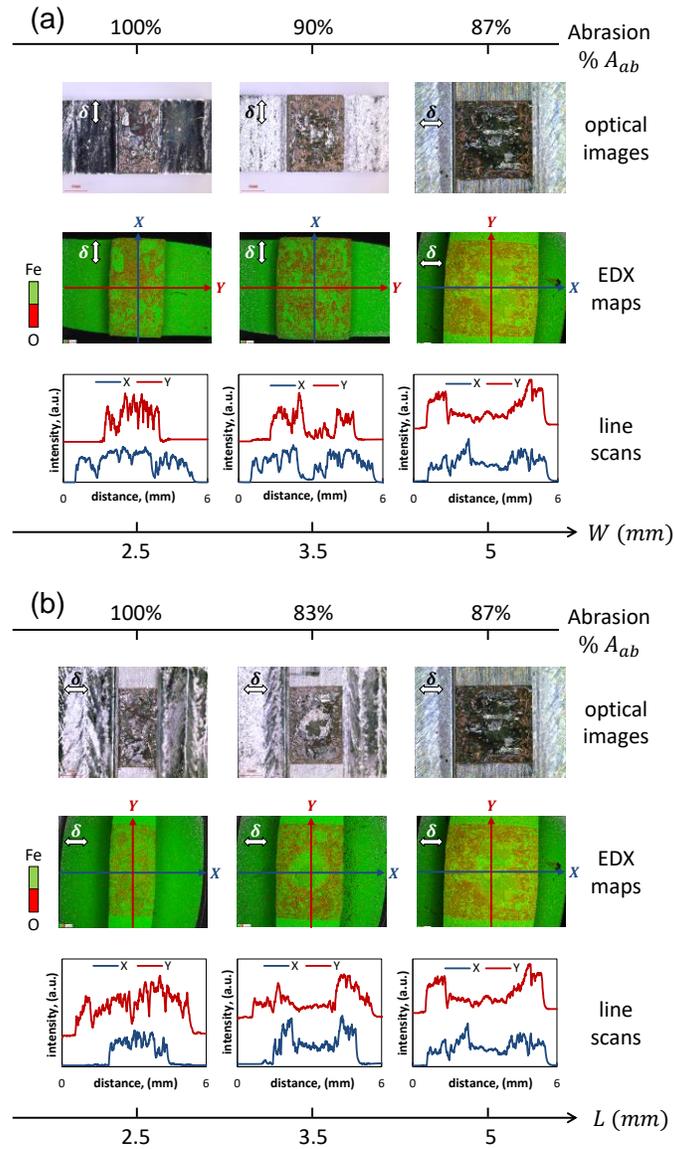

*Fig. 11. Optical images and EDX observations showing the recession of adhesive wear with the decrease of the contact area from 25 to 12.5 mm² (N=20000 cycles, p=100 MPa, $\delta_g$=±100 μm, and f=1 Hz).*

Such transition in wear regimes from a mixed abrasive-adhesive to a pure abrasive wear is confirmed in Fig.12a where the percentage of abrasion area increases from 87% at 25 mm² to 100% for contact areas less than or equal to 12.5 mm². Another significant remark is that no substantial difference is detected by varying the contact orientation with respect to the sliding direction for the same contact area. On the other hand, if the adhesive wear area decreases with the contact size, the oxygen distance remains constant and stabilizes around $d_O$ =1.37 mm (Fig. 12b) so that:

$$d_O = K_A \times d_{O,ref} \approx 1.37 \ mm \tag{17}$$

Where $K_A$ = 0.91 is an experimental correcting factor of the mean value compared to the reference test condition.

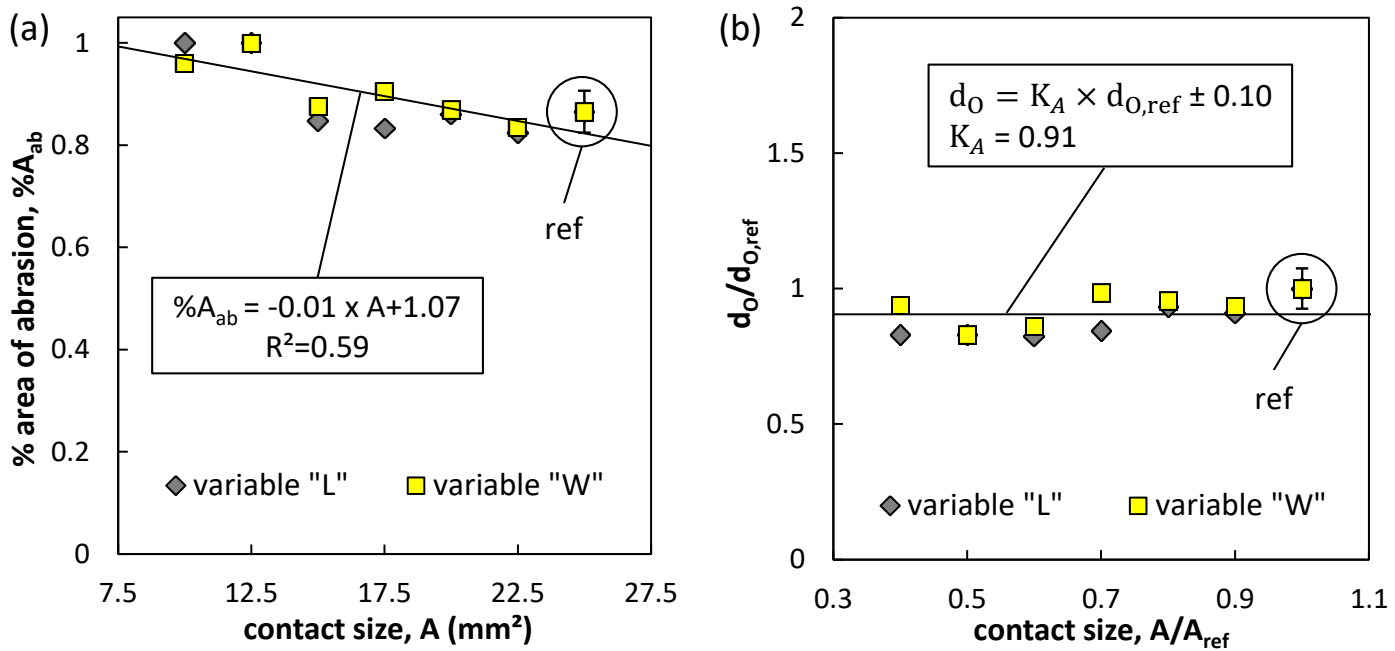

*Fig. 12.* Evolution of (a) the percentage of abrasive wear area "$\%A_{ab}$" and (b) the oxygen distance "$d_O$" as a function of contact size by varying either the longitudinal length (L) or the transverse width (W )(N=20000 cycles, p=100 MPa, $\delta_g$=±100 μm, and f=1 Hz with $A_{ref}$=5x5=25 mm² and $d_{O,ref}$ =1.51 mm).

### 3.1.6 A parametric "oxygen distance" modeling

The objective of the following part is to formalize the oxygen distance evolution as a function of the fretting parameters. The given experimental results suggest that the fretting cycles, contact area and sliding amplitude have a weak impact on oxygen distance $d_O$ at least for the studied interfaces. On the other hand, oxygen distance appeared to be significantly affected by the contact pressure and frequency. By combining Eqs. 13, 14, 15, 16 & 17, the oxygen distance "$d_O$" can be estimated using a very simple power law function:

$$d_O = d_{O,ref} \times (K_N \times K_\delta \times K_A \times K_f \times K_p) \times \left(\frac{f}{f_{ref}}\right)^{n_f} \times \left(\frac{p}{p_{ref}}\right)^{n_p}$$

$$\frac{d_O}{d_{O,ref}} = K_N \times K_\delta \times K_A \times K_f \times K_p \times \left(\frac{f}{f_{ref}}\right)^{n_f} \times \left(\frac{p}{p_{ref}}\right)^{n_p} \tag{18}$$

The correlation between the "K" coefficients leads to:

$$K_N \times K_\delta \times K_A \times K_f \times K_p \approx 1.0 \tag{19}$$

Hence, the predicted oxygen distance can be expressed using the simplified relationship:

$$d_O = d_{O,ref} \times \left(\frac{f}{f_{ref}}\right)^{n_f} \times \left(\frac{p}{p_{ref}}\right)^{n_p} \tag{20}$$

with $d_{O,ref} = 1.51\ mm$, $n_f = -0.22$, $n_p = -0.32$, $p_{ref} = 100\ MPa$, and $f_{ref} = 1\ Hz$

Fig. 13 compares the experimental and the predicted oxygen distance $d_O$ derived from Eq. 20. A good correlation is observed which suggests that the given formulation of $d_O$ value is able to formalize not only the transition from pure abrasive wear to a composite abrasive-adhesive fretting wear response but also to express the partition between the abrasive and adhesive wear domains inside the fretting scar.

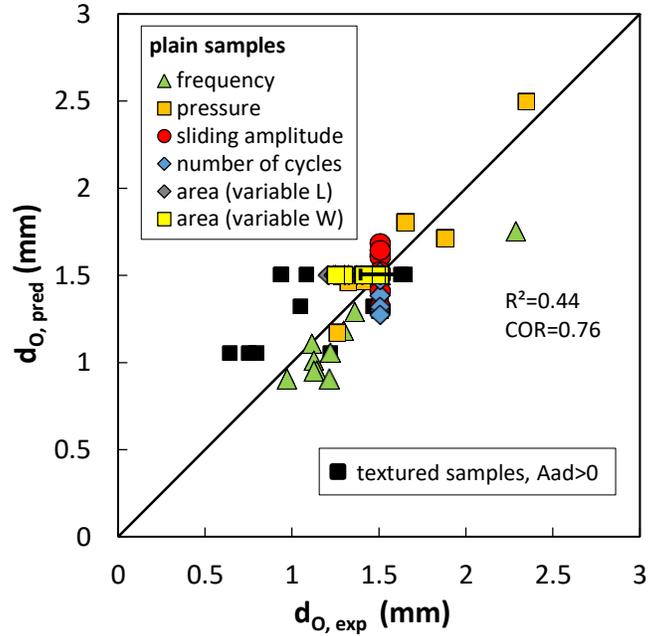

***Fig. 13.*** *Comparison between predicted and experimental oxygen distance using Eq. 20 applied on plain and textured samples at different loading conditions and contact geometries.*

3.2 Textured crossed flat-on-flat contact

The significance of predicting oxygen distance is that it permits not only detecting the nature of the wear regime (i.e. pure abrasive or mixed abrasive-adhesive wear), but also helps forecasting the transition between these regimes at different loading conditions. To check this aspect outside the calibration domain, textured bottom samples are tested. One interest of macro-textured surfaces is the possibility of investigating a broad range of "$d$" values (i.e. minimum distance from the contact center to the external air, (Eq. 5)) while keeping constant the longitudinal length "$L$".

3.2.1 Experimental strategy

The texturing strategy consists of machining rectangular grooves on the flat surface of the bottom samples (Fig. 14). The depth (D), the thickness ($t$), and the length ($L$) of the grooves are kept constant (1, 0.5 and 5 mm respectively). However, the spacing between the grooves is varied from 0.5 to 5 mm to adjust the rectangular pads' width "$W$". Note that the rectangular pads are systematically parallel the sliding direction (δ).

For "$W$" ranging from 0.5 to 5 mm, the contact pressure (p), the sliding amplitude ($δ_g$) and the frequency (f) are varied while maintaining a constant number of cycles N=20000. Three mean contact pressures p=50, 100 and 150 MPa were investigated along with three different sliding amplitudes $δ_g$=±50, ±100 and ±150 µm and three distinct frequencies f=0.5, 1 and 5 Hz. For each contact pressure value, the normal load (P) is adjusted to keep the desired contact pressure constant within textures' widths ($W$).

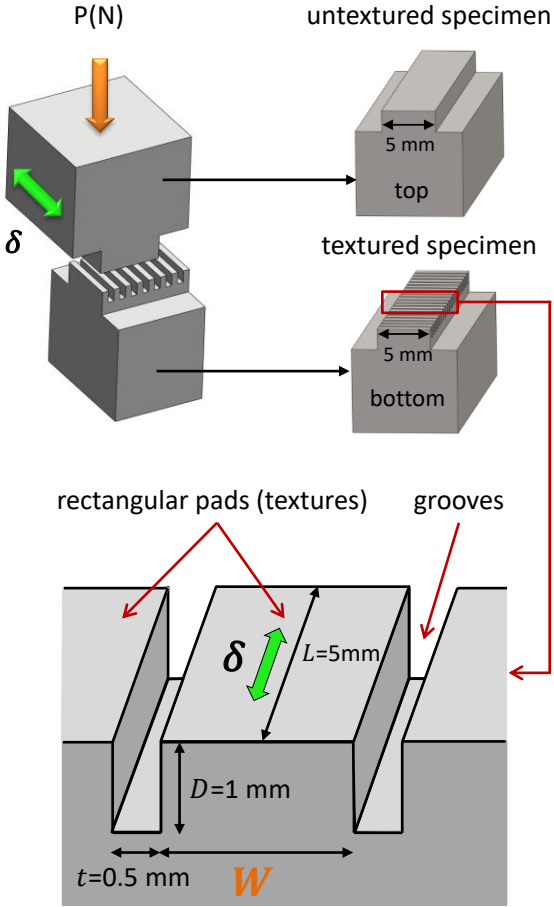

*Fig. 14.* Crossed flat-on-flat configuration with textured bottom sample.

### 3.2.2 Qualitative validation of the model

*3.2.2.1 Influence of the contact size and the sliding amplitude*

Fig. 15 shows the evolution of wear scars at different textures' width $W$ ranging from 0.5 to 5 mm (i.e. different areas, $A$) applying the reference test conditions. According to Eq. 20, the predicted oxygen distance marking the transition from pure abrasive to mixed abrasive-adhesive response is equal to $d_{O,pred} = d_{O,ref}$ =1.51 mm. After this consideration, when $W > 2.d_{O,pred} = 3\ mm$, a mixed regime is expected and below this value, we should observe pure abrasive wear. Careful observations of EDX maps reveal a mixed regime of abrasion and adhesion at $W$=5 mm and dominant abrasive wear for $W \leq 3\ mm$ which indirectly confirms the theoretical $d_O$ boundary. Note that like for mono plain contact configuration, the sliding amplitude plays a minor role as a constant transition is observed (Fig. 15 a to c).

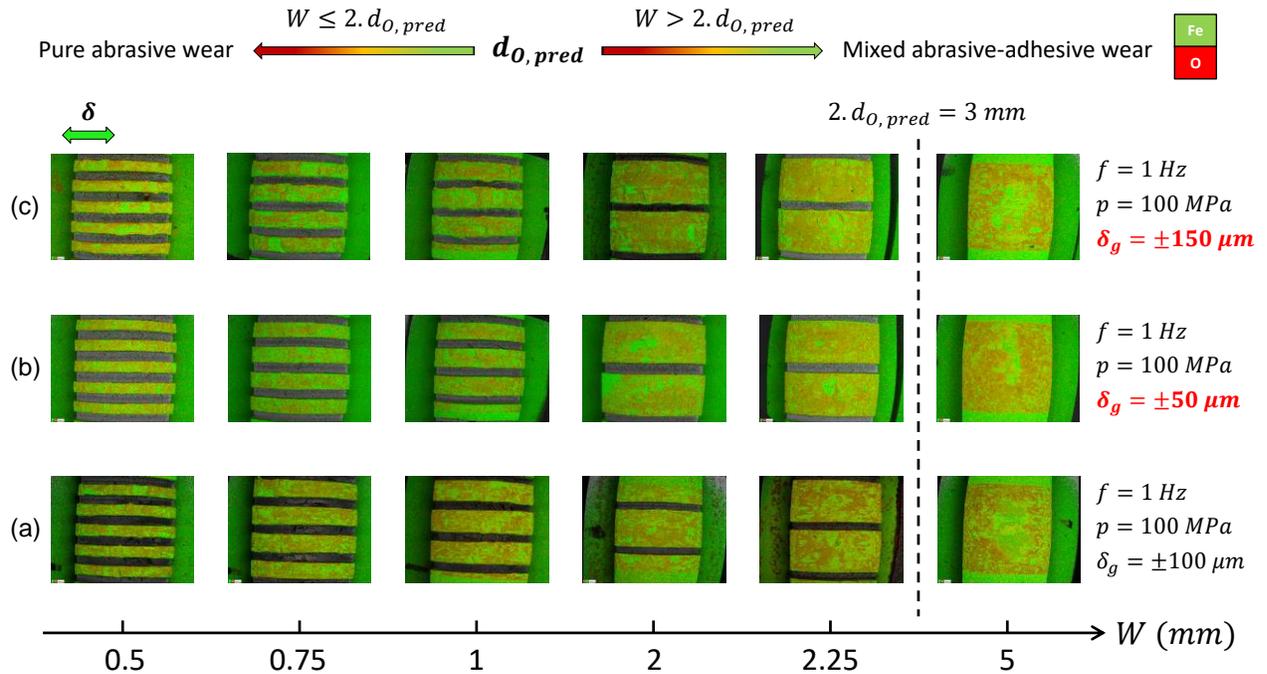

*Fig. 15. Prediction of the transition of wear regime by using the predicted threshold oxygen distance (Eq. 20) for: (a) variable contact area under reference loading conditions; (b) variable contact area at low sliding amplitude, δ_g=±50 μm; and (c) variable contact area at high sliding amplitude, δ_g=±150 μm.*

*3.2.2.2 Influence of the frequency and the contact pressure*

The capacity of predicting wear transitions of textured surfaces is also tested for variable contact pressure and frequency as displayed in Fig. 16. By reducing the contact pressure, it is expected to have transition of wear regimes at larger textures' width which is the case detected experimentally where the oxygen distance decreases by passing from p=50 to 150 MPa (Fig. 16 a & b). This transition is even clearer with the increase in frequency where a marked shift in the oxygen distance is noticed by comparing EDX maps at 0.5 and 5 Hz (Fig. 16 c & d).

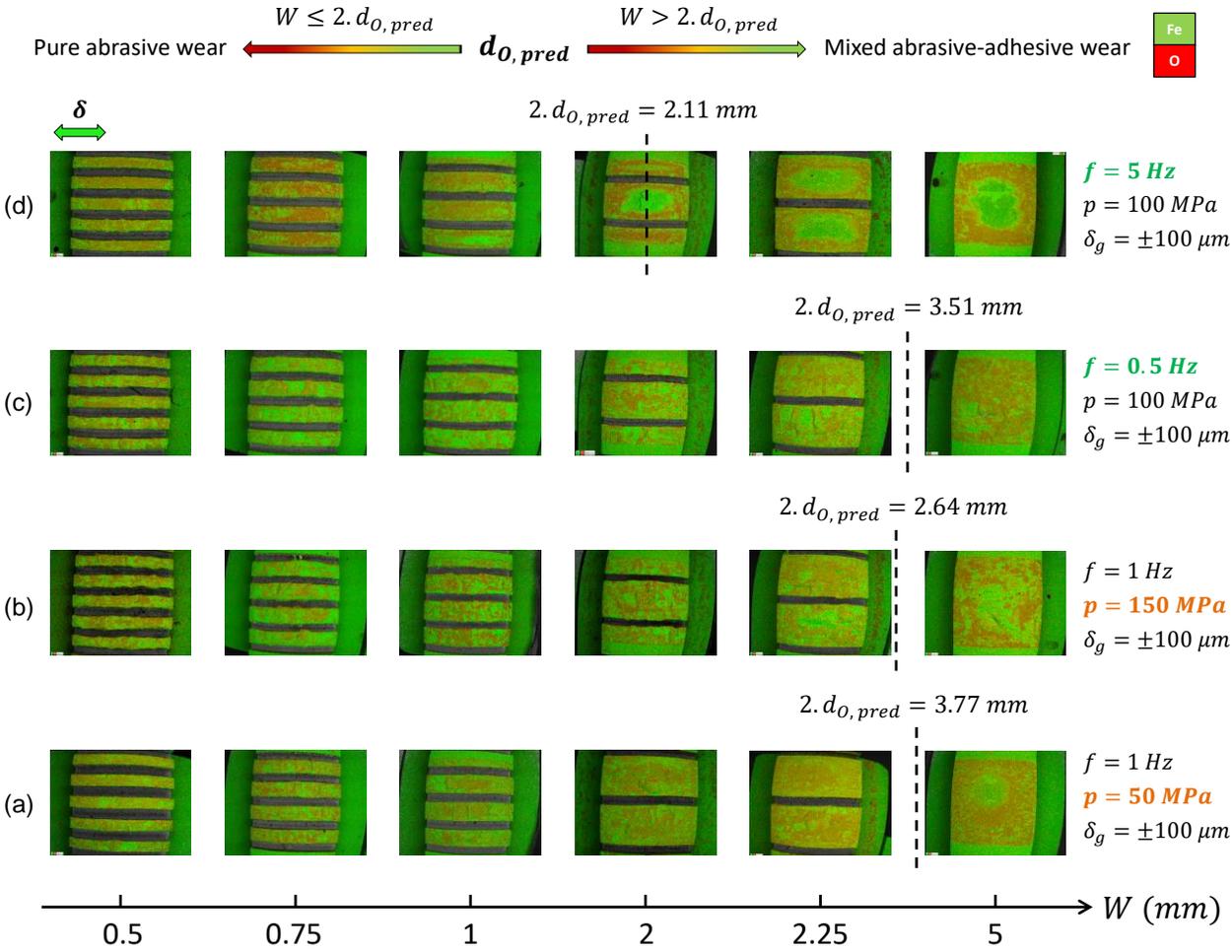

*Fig. 16. Prediction of the transition of wear regime by using the predicted threshold oxygen distance (Eq. 20) for: (a) variable contact area at low contact pressure, p=50 MPa; (b) variable contact area at high contact pressure, p=150 MPa; (c) variable*

*contact area at low frequency, f=0.5 Hz; and (d) variable contact area at high frequency, f=5 Hz.*

### 3.2.3 Validation of the parametric "oxygen distance" model

The parametric "oxygen distance" model (Eq. 20) is now considered to predict the extension of the $d_O$ oxygenation length scale for the studied textured samples. Note that when the lateral width is smaller than the full oxygenation condition (W < 2×$d_O$), so that the adhesion area is zero (i.e. $A_{ad}$=0), the corresponding data points are not considered in the analysis. Fig. 13 compiled both plain and textured crossed flat-on-flat contact results. A good correlation is still observed which confirms the stability of the approach to formalize the extension of abrasion domain within the fretting wear interface whatever the fretting loading condition and the contact morphology.

Now, assuming that the oxygen distance is equal in all directions and the inner adhesive wear is equivalent to a homothetic rectangular region (Fig. 17), the adhesive and abrasive wear areas "$A_{ad}$ & $A_{ab}$" as well as the proportions of the adhesive and abrasive wear areas "%$A_{ad}$ & %$A_{ab}$" can be expressed as a function of the $d_O$ formulation (Eq. 20) such that :

if $d = min\left(\frac{L}{2}, \frac{W}{2}\right) > d_O$,

$$A_{ad} = (W - 2.d_O).(L - 2.d_O) = W.L - 2.d_O(W + L) + 4.d_O^2$$

Hence,

$$A_{ad} = A - 2.d_{O,ref} \times \left(\frac{f}{f_{ref}}\right)^{n_f} \times \left(\frac{p}{p_{ref}}\right)^{n_p} \times (W + L) + 4.d_{O,ref}^2 \times \left(\frac{f}{f_{ref}}\right)^{2 \times n_f} \times \left(\frac{p}{p_{ref}}\right)^{2 \times n_p}$$

Which implies

$$A_{ab} = A - A_{ad} = 4.d_{O,ref}^2 \times \left(\frac{f}{f_{ref}}\right)^{2 \times n_f} \times \left(\frac{p}{p_{ref}}\right)^{2 \times n_p} - 2.d_{O,ref} \times \left(\frac{f}{f_{ref}}\right)^{n_f} \times \left(\frac{p}{p_{ref}}\right)^{n_p} \times (W + L)$$

if $d = min\left(\frac{L}{2}, \frac{W}{2}\right) \leq d_O$ then $A_{ad} = 0$ and $A_{ab} = A$  (21)

The relative proportions of abrasion (%A$_{ab}$) and adhesion (%A$_{ad}$) areas are expressed by Eq. 12.

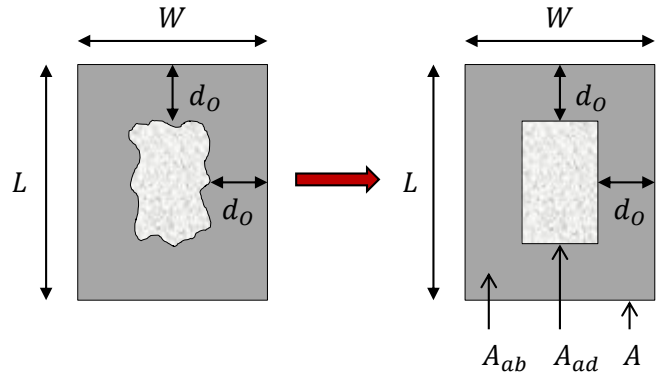

*Fig. 17. Illustration of the simplified partition abrasive and adhesive wear areas.*

The prediction of the corresponding effective adhesive (A$_{ad}$) and abrasive (A$_{ab}$) areas along with the relative abrasive wear areas (%$A_{ab}$) achieved using Eq. 20 and 21 are compared versus the experimental data in Fig. 18a, b & c respectively. By compiling the results of 104 tests (49 calibration tests with untextured plain bottom sample + 55 validation tests with textured bottom samples), very good correlations emerge especially for the prediction of abrasive and consequently adhesive wear areas where R²=0.98 and 0.86 correlation factors are observed respectively. Hence, it can be concluded that the given formulation of $d_O$ (Eq. 20) and the consequent expressions of the abrasive and adhesive wear areas "%$A_{ab}$" and "%$A_{ad}$" are very reliable and may be considered to formalize the fretting scar morphology at least for the studied crossed flat contact configurations inducing constant contact areas and nearly constant mean pressure during the test.

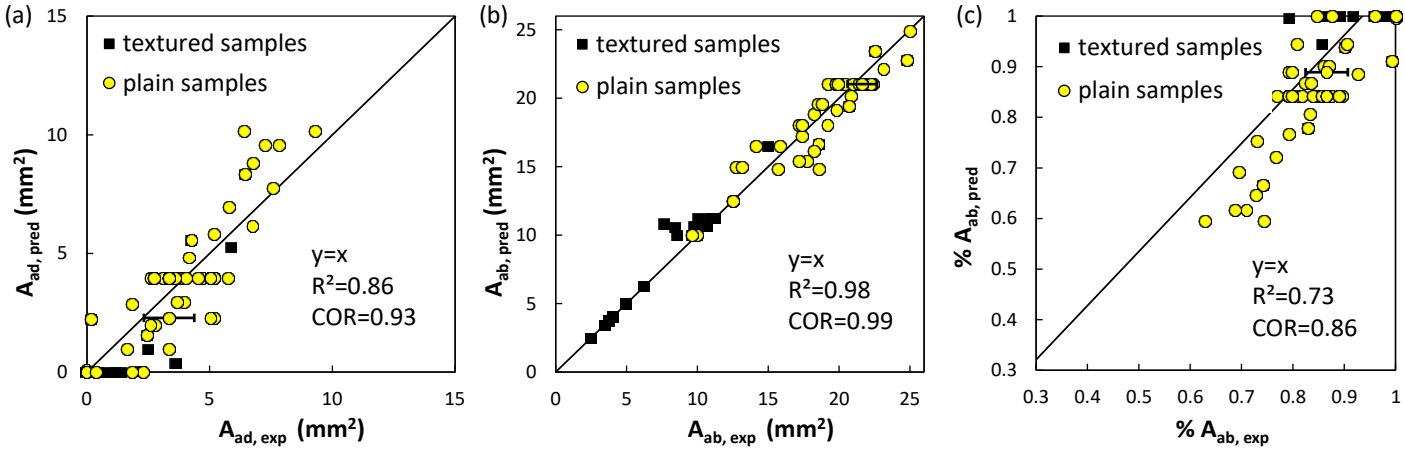

***Fig. 18.*** *Correlation between the experimental and the predicted results (Eq. 20 to 21) for (a) the adhesion area "$A_{ad}$"; (b) the abrasion area "$A_{ab}$" and (c) the relative proportion of abrasion area "$\%A_{ab}$" by compiling all the tests for plain and textured samples.*

## 4. Discussion

The former investigation allows us to quantify and formalize the relative extension of abrasive and adhesive wear areas within a flat-on-flat fretting interface. A simple parametric formulation was extracted taking into account most of the fretting loading parameters such as the sliding frequency, the mean contact pressure, the sliding amplitude, the test duration and also the contact size and contact morphology including complex macro-textured interfaces. There is however a crucial interest to interpret in a physical point of view the relative influence of each of these parameters. To do this, we consider the IOC concept previously introduced in [20].

4.1 Frequency effect

The asymptotic decreasing of $d_O$ versus frequency (Fig. 7) is consistent with the "contact oxygenation" concept [20] (Fig. 19). Indeed by increasing the sliding frequency, the friction power density inputted in the interface rises which promotes a faster fresh metal exposition and therefore a faster consumption of available dioxygen molecule within the fretting interface and therefore a reduction of $d_O$. Moreover it reduces the time allowed to the dioxygen molecule to diffuse inward the inner part of the fretting contact. These combined effects tend to decrease the interface dioxygen partial pressure which can explain the reduction of the well oxygenated lateral abrasive domain $d_O$ and the corresponding extension of the inner oxygen-deprived adhesive wear area. However, it is difficult to determine which of the above effects is preponderant in the absence of available experimental tools indispensable for quantifying the dioxygen diffusion, rate of di-oxygen consumption, etc.,..

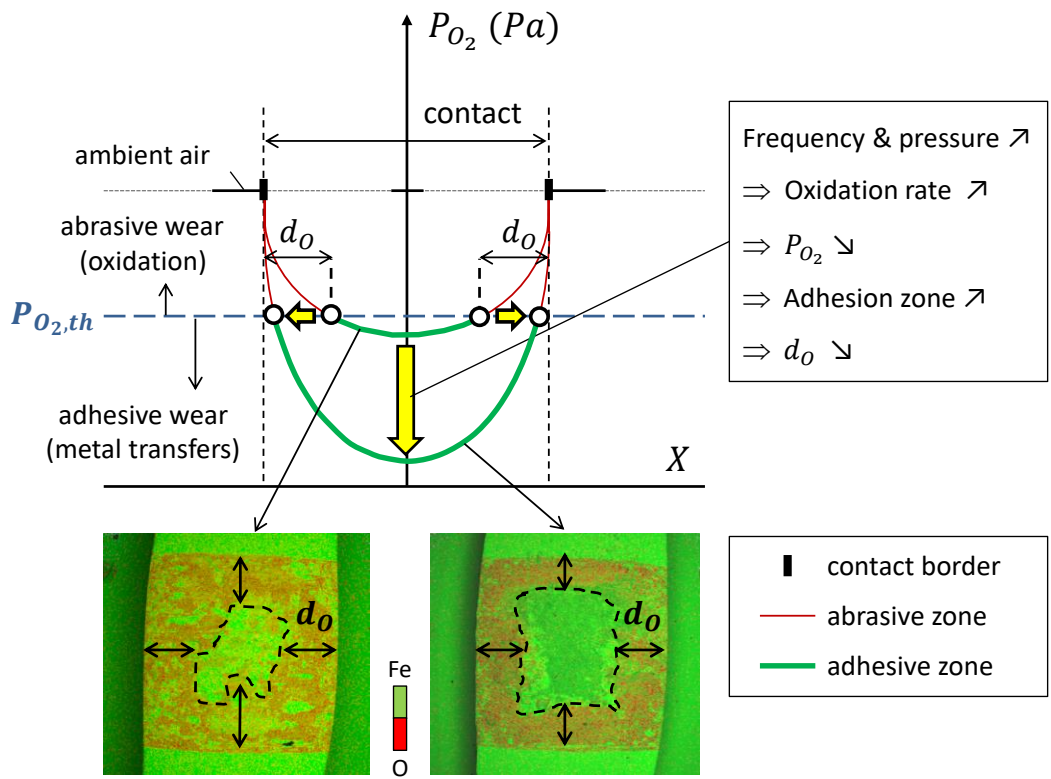

***Fig. 19.*** *Illustration of the contact oxygenation concept to interpret the effect of contact pressure and sliding frequency regarding the evolution of $d_O$ and the related partition between adhesive and abrasive areas.*

4.2 Contact pressure effect

Like frequency, an increase of the contact pressure promotes a rising of the friction power density and consequently a faster consumption of the dioxygen molecule available within the fretting interface. Moreover, an increase of contact pressure, by compacting the surface roughness and the debris layer, tends to reduce the diffusion of dioxygen molecules from the open air toward the inner part of the contact. These two combined effects drastically reduce the interfacial di-oxygen partial pressure (Fig. 19) which indirectly explains the fast reduction of $d_O$ versus contact pressure observed in Fig. 8. (Eq. 14). It is noteworthy to mention the larger $n_p$ pressure exponent compared to the corresponding frequency value (i.e. $|n_p|/|n_f| \approx 1.5$). This suggests a dominant effect of contact pressure regarding the partition between abrasive and adhesive wear compared to the frequency effect. This dominant effect might come from the fact that increasing the contact pressure causes compaction of the debris bed which reduces the porosity and hence the permeability of the latter. This particular effect might not be significant when the frequency is increased, a possible reason why it has smaller exponent compared to the contact pressure.

4.3 Test duration effect

It is interesting to underline the very fast convergence of the interface toward a stabilized partition between abrasive and adhesive wear zone (Fig. 9). The $d_O$ length scale stabilized before 5000 fretting cycle and then remained constant. This suggests a constant IOP profile (i.e. boundary between well-oxygenated abrasive domains and under-oxygenated adhesive domains) whatever the surface wear extension. This tendency is coherent with the contact oxygenation concept in the sense that the studied crossed flat configuration imposes constant contact area, constant mean pressure and therefore constant friction work density. A different behavior should be observed if a Hertzian contact configuration was investigated. Indeed, a Hertzian contact, which displays a huge extension of the worn contact area, is expected to induce a significant increase of the $d_O$ length scale with the fretting cycles and potentially an evolution from a composite adhesive-abrasive wear response toward a pure abrasive wear process.

Unfortunately it was not possible in the frame of this investigation to study different contact geometries. The given stable evolution of $d_O$ needs however to be considered with caution for very long test conditions where the progressive surface wear could eventually modify the dioxygen diffusion rate within the interface and therefore change the partition between abrasive and adhesive wear domains.

4.4 Sliding amplitude effect

It is surprising to notice a quasi-constant evolution of the $d_O$ length scale versus the studied ± 25 to ± 200 µm sliding amplitude range (Fig. 10). Longer sliding amplitudes increase the contact exposure to the outer environment (e= $δ_g$/a) causing recession of adhesive zone [6,29–31]. This was observed in many studies especially for very large fretting sliding amplitudes equivalent for instance to the reciprocating sliding case (e≥1). However in the present situation the sliding amplitude remains smaller than 8% compared to the half contact width. This implies that most of the interface remains hidden from the outer environment which may explain the rather small influence of the studied sliding amplitude regarding the $d_O$ parameter.

To interpret such constant evolution of $d_O$, the "contact oxygenation" concept will be considered again. An increase in the sliding amplitude causes a proportional increase in the friction power density (i.e. µ.p.v with µ being the friction coefficient) and therefore acceleration in the dioxygen reaction rate with fresh metal surfaces exposed by the tribological processes. This implies a decreasing of the IOP profile and consequently a reduction of the abrasive domain. The fact that $d_O$ remains constant suggests that another mechanism is counterbalancing such effect. One hypothesis could be that an increase of the sliding amplitude also enhances debris flow and consequently the open access between interlocked asperities favoring in turn the diffusion of dioxygen molecules inward the fretted interface. These two opposite effects (increase of dioxygen consumption rate and easier diffusion of dioxygen molecule from outer ambient air) could explain the constant evolution of $d_O$ with the sliding amplitude at least for the studied condition. However, proving experimentally the validity of these hypotheses remains a scientific challenge.

It is clear that the application of longer sliding amplitude may shift the balance between the two processes and consequently modify the oxygen distance parameter extending progressively the abrasive wear area when the sliding amplitudes approach the reciprocating sliding condition.

4.5 Contact size and sliding orientation

Interfacial di-Oxygen Concentration concept (IOC) [20] (i.e. equivalent to the given IOP partial pressure description), suggests that the extension of $d_O$ (i.e. lateral abrasive wear domain) is a function of the balance between the dioxygen molecule diffusion rate from the external contact borders and the rate of oxygen consumption related to the oxidation process of fresh metal surface and metal debris induced by the friction work. Assuming constant friction work density condition (i.e. constant pressure and frequency for the studied flat-on-flat interface), contact size is not involved in such description and therefore $d_O$ should be independent of contact area. This conclusion is confirmed experimentally in Fig.12b. Obviously, when the minimum distance between the contact border and the contact center becomes smaller than $d_O$, the interface shifts towards a pure abrasive wear interface: $d < d_O \Rightarrow$ % abrasive area=100%. Alternatively if $d_O$ remains constant, an increase of the contact size tends to increase the inner adhesive area, promoting a reduction of the relative abrasive area (%$A_{ab}$) as confirmed in Fig. 12a.

The isotropic distribution of $d_O$ around the fretted scar whatever the contact size and orientation (i.e. $d_{O,L} \approx d_{O,W}$) can be explained by the isotropic diffusion of dioxygen molecules (i.e. $d_{O,X} \approx d_{O,Y}$ in Fig. 6b). Such isotropic response is not so obvious. Indeed, the studied crossed flat contact allows homogeneous friction power density dissipation over the whole fretted interface. Therefore, homogeneous dioxygen consumption rate is expected. However, it could be assumed that the longitudinal abrasive grooves along the sliding direction will allow easier oxygen diffusion inward the center of the fretted interface. In contrast, focusing on the transverse direction, dioxygen molecules need to overpass the perpendicular roughness barriers related to the longitudinal grooves. From this topographic consideration, it might be expected larger $d_{O,X}$ than $d_{O,Y}$ values.

The fact that similar $d_{O,X}$ and $d_{O,Y}$ distances are observed suggests that isotropic diffusion conditions, (i.e. independent of the sliding and roughness directions) are ensured within the fretted contact. This result could be explained by assuming that the interfacial dioxygen diffusion process is mainly driven the isotropic 10 µm thick porous debris layer than by 1 µm anisotropic fluctuation of fretting scar roughness.

4.6 Synthesis

To conclude, it can be stated that the proposed IOC concept and the related di-oxygen partial pressure distribution approach appear as a pertinent strategy to explain the typical abrasive-adhesive composite structure observed in large fretting wear interface. Using such simple concept, it is possible to explain the relative influence of frequency, pressure, loading cycles, sliding amplitude and contact size effects regarding the extension of the inner adhesive zone in fretting contact. A basic formulation is also introduced allowing the formalization of the $d_O$ length scale (i.e. surrounding "well oxygenated" lateral band of the fretted interface) as well as the corresponding adhesive and abrasive areas. The very good correlation with the experiments confirms the proposal and was even able to formalize the fretting wear response of complex macro textured interfaces. This formalization of the composite adhesive-abrasive fretting wear interface will permit a better description of the global wear rate such that different wear rates should be expected in the distinct adhesion and abrasion zones. It will also permit a better prediction of the fretting cracking process observed in large and low pressure interfaces where adhesive interface can localize contact stressing favoring fretting crack nucleation process [7,32]. Future works will be undertaken to investigate the influence of the contact geometry studying for instance various Hertzian contact configurations. Another aspect will concern the investigation of various materials like bronze, aluminum and titanium alloys displaying various oxidation rates and adhesive wear properties. Hence by tuning the balance between di-oxygen diffusion and di-oxygen depletion it will be possible to establish a more physical formulation of the $d_O$ length scale. However, the main objective of the current development consists of the establishment of an explicit formulation of the di-oxygen partial pressure profile thus to provide a robust prediction of $d_O$ and the related partition between adhesive and abrasive wear zones in fretting interfaces. This approach, though being fundamental, might be profitable for industrial practice as it allows better prediction of wear kinetics and wear depth by taking into account mixed abrasive-adhesive wear which is highly encountered in many industrial applications especially those operating in oxygen-starved environments like large press-fitted assemblies where seizure may occur. Note that seizure phenomenon can favor crack nucleation and fretting fatigue failures [33]. Therefore, the prediction of adhesive wear zone is a key issue to establish safe fretting fatigue interface.

## 5. Conclusion

This paper aims at investigating the evolution and transition of wear mechanisms (abrasion and adhesion) for a 34NiCrMo16 dry flat-on-flat interface. This is achieved by introducing the "oxygen distance, $d_O$" defined as the distance between the contact extremities and the limits of adhesion domain. Then, a multi-scale experimental strategy is applied where the number of cycles, contact pressure, sliding amplitude, frequency, and contact area and geometry are varied.

Results showed that oxygen distance depends mainly on pressure and frequency. This is due to higher dissipated friction energy which leads to high plastic deformations and decrease in the time given for the oxidation to take place (in case of frequency). On the other hand, oxygen distance was shown to be stable with the number of cycles, sliding amplitude and contact area. Taking these results into account, an oxygen distance prediction model was proposed which depends on pressure and frequency. This model was validated by using textured samples with variable textures' width. In addition, to predicting abrasion area, this model helped detect the transition of wear mechanisms from full abrasion to a mixed abrasive-adhesive wear under coupled loading conditions.

**Acknowledgement**

The authors gratefully thank the French National Research Agency (ANR, France) and the Ecole Centrale de Lyon (ECL, France) for financially supporting this research project (ANR-16-CE08-0016). Besides, thanks are extended to Julie Laporte and Jean-Michel Vernet for their technical help.

**Statement of Originality**

I, Siegfried Fouvry, the corresponding author of the submitted paper state that this work has not been submitted elsewhere.

This paper develops an original experimental analysis to formalize the transition between abrasive and adhesive wear behaviour of a low alloyed steel interface subjected to gross slip fretting wear. The new innovative aspects are the following:

- Development of an original crossed flat-on-flat fretting wear experiment.

- Quantification of the partition between abrasive and adhesive wear zones in the fretting scar using the so-called $d_O$ "oxygen distance" parameter defined as the averaged width of the external abrasive wear corona observed in the fretting scar.

- The analysis quantifies the evolution of $d_O$ as a function of the sliding frequency, contact pressure, fretting cycles, sliding amplitude but also contact area and siding orientation.

- A power law function is introduced to formalize the evolution of $d_O$ as a function of the latter loading variables.

- A very good correlation is observed between the experiments and the model.

- The model is even able to predict the fretting wear response of complex macro-textured interfaces.

- These different results are discussed regarding the "Contact Oxygenation Concept". Using this approach it is possible to explain why $d_O$ decreases with pressure and frequency but remains constant versus sliding amplitude, fretting cycles and contact area.